\documentclass[preprint2]{aastex}

\def\fun#1#2{\lower3.6pt\vbox{\baselineskip0pt\lineskip.9pt
  \ialign{$\mathsurround=0pt#1\hfil##\hfil$\crcr#2\crcr\sim\crcr}}}

\begin{document}

\title{Supermassive Black Holes}

\author{Laura Ferrarese and David Merritt}
\affil{\phantom{aaaaa}}

\bigskip\bigskip

\noindent
Of all the legacies of Einstein's general theory of relativity, none is more fascinating than black 
holes. While we now take their existence almost for granted, black holes were viewed for much of 
the 20th century as mathematical curiosities with no counterparts in nature. Einstein himself never 
believed in black holes and wrote two papers in which he argued against their existence. 
Einstein's resistance to the idea is understandable. 
Like most physicists of his day, he found it hard to 
believe that nature could permit the formation of objects as extreme as black holes.
Indeed, the gravitational fields of black holes are strong enough to prevent light from escaping, 
and even distort space and the flow of time around them.

The modern view -- that black holes are the unavoidable end result of the evolution of massive 
stars -- arose from the work of 
Subrahmanyan Chandrasekhar, Lev Landau, Robert Oppenheimer and others in the first half of the 
20th century. However it was not until the discovery in 1963 of extremely luminous distant objects 
called quasars that the existence of black holes was generally acknowledged. What is more, black 
holes appeared to exist on a scale far larger than anyone had anticipated.

Quasi-stellar objects or quasars belong to a class of galaxies known as active galactic nuclei. 
What makes these galaxies ``active'' is the emission of staggering amounts of energy from their 
nuclei. Moreover, the luminosities of active galactic nuclei fluctuate on very short time scales -- 
within days or sometimes even minutes. The time variation sets an upper limit on the size of the 
emitting region. For this reason we know that the emitting regions of active galactic nuclei are only 
light-minutes or light-days across, making them less than one ten-millionth the size of the galaxy in 
which they sit.
Astronomers were faced with a daunting task: to explain how a luminosity hundreds of times that 
of an entire galaxy could be emitted from a volume billions of times smaller. Of all proposed 
explanations, only one survived close scrutiny: the release of gravitational energy by matter falling 
towards a black hole. 
Even using an energy source as efficient as gravity, the black holes
in active galactic nuclei would need to be enormous --
millions or even billions of times more massive than the Sun -- in order
to produce the luminosities of quasars. 
To distinguish these black holes from the stellar-mass black holes left behind by supernova 
explosions, the term ``supermassive black hole'' was coined.

\bigskip
\noindent
{\bf Where have all the quasars gone?}

\noindent
For nearly three decades after quasars were discovered, supermassive black holes
continued to be viewed as exotic phenomena and their existence was accepted only out of 
necessity. However, by the late 1980s a major crisis was brewing. Surveys with optical telescopes 
had shown that the number of quasars per unit volume is not constant with time. By studying the 
redshift of the light emitted by the quasar on its journey to the Earth, astronomers found that the 
number density of quasars peaked when the universe was only about two and a half billion years 
old and has been declining steadily ever since.

\begin{figure}
\epsscale{0.9}
\plotone{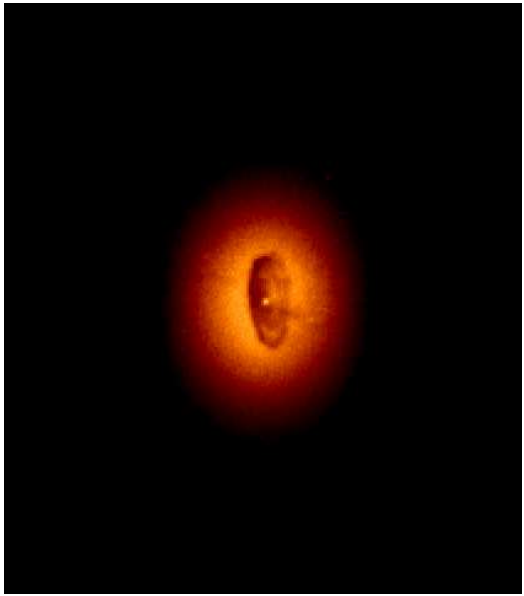}
\caption{{\bf Fuelling the monster.}
A Hubble Space Telescope image of NGC 4261, an elliptical galaxy 100 million light-years away. 
The dark outline is produced by a disc of gas and dust measuring about 780 light years across, 
which obscures the stellar light from the far side of the galaxy. The disc, which is slightly inclined 
with respect to the line of sight, is believed to be the ultimate source of fuel for the central 
supermassive black hole: the gravitational energy released by the gas just before plunging into the 
event horizon is seen as the bright spot of light at the centre of the disc. By modelling the velocity 
of the gas in the innermost parts of the disc, we believe the central black hole to be 50 million 
times the mass of the Sun.
}
\end{figure}

The reason for this evolution remains one of the great unsolved mysteries of modern astrophysics. 
But it also presents astronomers with an additional challenge.
Many of the quasars with large redshifts simply 
disappear at lower redshifts. Indeed, of the quasars that 
populated the skies almost 10 billion years ago, only one in 500 can be identified today -- but 
we know of no way to destroy the supermassive black holes that powered the quasar activity. The unavoidable 
conclusion is that the local universe is filled with ``dead'' quasars, supermassive black holes that 
have exhausted the fuel supply that made the quasars shine so brightly 10 billion years ago.

Where are these dead quasars? A reasonable place to look is at the centres of active galactic 
nuclei. But while these active galactic nuclei almost certainly do contain supermassive black holes, there are far too 
few of them -- only about 1\% of all galaxies are active -- to account for all of the supermassive black holes that once 
powered the quasars. By the early 1990s, astronomers were faced with the prospect 
that a supermassive black hole might have to be located at the centre of every galaxy, 
making them as fundamental a 
component of galactic structure as stars. In other words, perhaps every galaxy was once shone as 
brightly as a quasar.

This idea -- though natural enough -- did not come easily, since most galaxies show no evidence 
for the emissions associated with a central supermassive black hole. In the words of Andy Fabian at Cambridge 
University and Claude Canizares at the Massachusetts Institute of Technology (MIT), ``starving the 
monster'' is not an easy task. There is more than enough gas and dust in any galaxy to make a 
central supermassive black hole shine as brightly as a quasar. It was a challenge for astronomers in the 1990s to 
prove that these objects actually reside in the nuclei of every galaxy.

\bigskip
\noindent
{\bf The first detection}

\noindent
By their nature, black holes cannot be observed directly. However, we can infer their existence 
from the motions of surrounding matter. To be perfectly rigorous, one needs to probe the region 
close to the ``event horizon'' -- the point beyond which matter cannot escape. Near this 
boundary the orbital velocity around the black hole is close to the speed of light.

Tremendous strides in this direction have recently been taken, thanks to the vast improvement in 
energy resolution brought by the most recent generation of X-ray telescopes. In 1995, when the 
ASCA satellite telescope was pointed toward the nearby active galactic nucleus MCG-6-30-15, a 
group of Japanese and British astronomers led by Yasuo Tanaka of the Institute for Space and 
Astronautical Science in Kanagawa noticed that the X-ray spectrum of the galaxy was dominated 
by an emission line due to iron. The presence of the line emission -- a characteristic of hot gas -- 
was not surprising, but its appearance was. The line was expected to be quite thin, or 
``monochromatic'', and symmetric. However, it was observed to be extremely broad and 
significantly skewed. Within two months, Andy Fabian at Cambridge University and collaborators 
had demonstrated that the peculiar shape of the iron emission was due to relativistic effects, which 
are only expected near the event horizon of a supermassive black hole.

The X-ray line of iron continues to be a powerful probe of the immediate environment of the central 
supermassive black hole in MCG-6-30-15 and a handful of other galaxies, as well as a unique tool for testing our 
knowledge of the properties of space and time in strong gravitational fields. However, astronomers 
will have to wait until the launch of Constellation-X, a revolutionary X-ray satellite expected to 
begin operations at the end of the decade, to use the X-ray line of iron to derive the most critical 
piece of information about a supermassive black hole: its mass.

Fortunately, we can take a different approach. The gravitational field of a supermassive black hole is strong enough to 
imprint a characteristic signature on the motion of surrounding matter even at distances that are 
millions times greater than the event horizon. Stars, gas and dust moving around a black hole -- or 
any compact object -- have orbital velocities that follow the same laws discovered
by Johannes Kepler in the 17th century for the solar system.
Moreover, the mass of the compact object is easily computed once this Keplerian rotation has 
been mapped. These arguments have been applied in spectacular fashion to measure the mass of 
the supermassive black hole in the core of our own galaxy, the Milky Way, and in the nearby galaxy NGC 4258.

Sufficient radiation is produced in the nucleus of galaxy NGC 4258 to excite water molecules in the 
molecular clouds, which leads to strong stimulated emission at radio wavelengths. 
These so-called water masers can be 
studied with very high spatial and velocity resolution using the interferometric techniques 
implemented in the Very Long Baseline Array (VLBA). In 1994, Makoto Miyoshi of the Mizusawa 
Astrogeodynamics Observatory and collaborators reported that the maser clouds traced a very 
thin disc, which made their dynamics easy to interpret. They found that the motion of 
the clouds within the disc followed Kepler's law to within one part in 100, reaching a velocity of 
1100 km s$^{-1}$ at a distance of 0.5 light years from the centre! Only by assuming that the nucleus of 
NGC 4258 -- already known to be an active galaxy -- hosts a central body with a mass 40 million 
times greater than the Sun could these observations be explained.

\begin{figure}
\epsscale{0.9}
\plotone{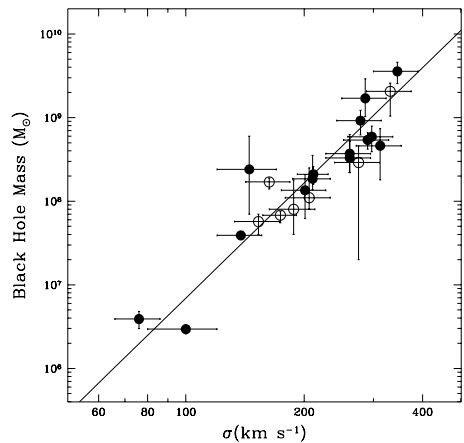}
\caption{{\bf Perfect correlation.}
The mass of supermassive black holes as a function of the velocity dispersion of the stars in the 
host galaxy. Filled circles are from published data; open circles are based on unpublished 
analyses. The scatter of points about the best-fit line is fully consistent with that expected on the 
basis of measurement errors alone (shown by the error bars), implying that the underlying 
correlation is essentially perfect.
}
\end{figure}

\bigskip
\noindent
{\bf Black holes on our doorstep}

\noindent
Perhaps even more remarkable is the case of the supermassive black hole at the centre of the Milky Way. The galactic 
centre has long been known to host a powerful radio source, called Sagittarius A* (SgrA*) that is at rest, 
indicating that it must be very massive. We also know that it is less than three billion kilometres 
across, which is similar in size to the orbit of Saturn. However, it took almost 70 years following the 
discovery of SgrA* for ground-based telescopes and analysis techniques to demonstrate that SgrA* 
is in fact a supermassive black hole.

For the past eight years, two groups -- one led by Andrea Ghez at the University of California at 
Los Angeles, the other by Reinhard Genzel at the Max Planck Institute in Garching -- have 
painstakingly monitored over 200 stars within 3 light years of SgrA*. The stellar motions have 
been reconstructed by combining the projected motion on the plane of the sky (referred to as 
``proper motion'') with the velocity along the line of sight. This latter component was measured from 
the Doppler shifts of absorption lines in the stellar spectra.

Ghez and Genzel's data revealed the unmistakable fingerprint of a supermassive black hole: stars closer to SgrA* 
move faster than stars farther away in the exact ratio predicted by Kepler's law. Stars 
only a few light-days away from the source move at fantastic speeds, in excess of 1000 km s$^{-1}$. 
Such velocities can only be maintained if SgrA* is roughly 3 million times more massive than 
our Sun.

The galactic centre is 100 times closer than the next large galaxy, Andromeda, and 2000 times 
closer than the nearby association of galaxies, the Virgo cluster. In no other galaxy do we have the 
opportunity to study the dynamics of individual stars orbiting a central supermassive black hole in such exquisite detail. 
To make matters worse, water masers like the one that populates the nucleus of NGC 4258 are 
very rare, and even more rarely are they organized in simple dynamical structures that can be 
easily interpreted.

But we have not yet reached the end of the road. The key is understanding what makes the 
observations in NGC 4258 and the Milky Way so successful. In both cases, the data probe regions 
in which the stellar or gas motions are completely dominated by the gravitational potential of the 
supermassive black hole. If we were to look further from the centre of these galaxies, we would find 
that the motion of the 
stars and gas clouds is influenced more by the spatial distribution of all the other nearby stars 
than by the central black hole. For instance, no information about the supermassive black hole at the galactic centre 
can be learned from the motion of the Sun.

In this regard, it is useful to define the ``black-hole sphere of influence'' as the region of space 
within which the black hole gravitational potential dominates that of the surrounding stars. Ghez 
and Genzel's stars, and Miyoshi's water masers, are buried well within the sphere of influence of 
the supermassive black holes in the Milky Way and in NGC 4258. With a factor 10 increase in spatial resolution 
compared with ground based telescope, the Hubble Space Telescope has now allowed us to 
glimpse the sphere of influence of the supermassive black holes at the centre of almost two dozen galaxies (figure 1). 
It has also recorded the most massive black hole detected to date: 3.5 billion solar masses, 
belonging to the central galaxy in the Virgo cluster, M87.

While painting a less detailed a picture of these galaxies than ground-based radio telescopes do 
for the Milky Way and NGC 4258, Hubble observations have one major advantage: they allow us to 
study supermassive black holes in a large number of galaxies. For the first time, we can produce an accurate census 
of supermassive black holes in the nearby universe. In fact, a supermassive black hole has been found in all the galaxies for which Hubble 
can resolve the sphere of influence around the black hole. The emerging scenario is one in which 
supermassive black holes are a fundamental constituent of all galaxies.

\bigskip
\noindent
{\bf Black hole demography}

\noindent
Confident of the existence of supermassive black holes, we can begin to ask more fundamental 
questions about them. How are black holes related to their host galaxies? How did they form? 
And what role do they play in galaxy evolution?

The answer to the first question is starting to emerge. Two years ago, a remarkable correlation 
between supermassive black holes and the properties of their host galaxies was discovered by the
authors of this article (see Ferrarese \& Merritt 2000 in further reading, and figure 2), 
and independently by Karl Gebhardt, now at the University of Texas in Austin, and co-workers. 
It turns out that the mass of a black hole can be predicted with remarkable accuracy by 
measuring a single number -- the stellar velocity dispersion, $\sigma$, in the host galaxy. While 
the orbits of single stars are fairly regular, they move in all directions like bees in a hive. The 
velocity dispersion is just a measure of the typical velocity of stars moving through a given point. 

What is so surprising about this relation, aside from its remarkable precision, 
is that the measurements need not be restricted to stars that are directly influenced by the 
black hole's gravitational field. In other words, 
supermassive black holes appear to ``know'' about the motions of stars that are far too distant to feel their gravity.

The origin of this relation is still being debated by theorists. But whatever its ultimate meaning, the 
relation is an extremely valuable tool because it links something that is difficult to measure (the 
mass of a supermassive black hole) to something that is easy to measure (the stellar velocity dispersion far from the 
supermassive black hole). It is therefore possible to determine the masses of supermassive black holes in a very large sample of galaxies, 
much larger than the sample for which the techniques described previously can be applied. When 
we did this, we found that about 0.1\% of a galaxy's luminous mass is associated with the supermassive black hole, 
and that the density of supermassive black holes in the local universe agrees remarkably well with the density inferred 
from observations of quasars. For the first time, the supermassive black holes that powered the distant quasars have 
been fully accounted for.

\bigskip
\noindent
{\bf Merging black holes}

\noindent
One of the most fascinating connections between supermassive black holes and galaxies was made by
Mitch Begelman at the University of Colorado, Roger Blandford at the California Institute of 
Technology and Martin Rees of Cambridge University. 
Following an earlier idea by Alar Toomre of MIT, they asked what would happen if two galaxies
containing such objects were to collide and merge to form a single galaxy.

Current models of galaxy formation suggest that most large galaxies have 
experienced at least one major merger during their lifetime (see ``How are Galaxies Made?'' 
{\it Physics World} May 1999 pp25--30). In a galactic merger, the supermassive black holes at the centres of the two 
galaxies would sink rapidly to the centre of the merged system through a process called 
``dynamical friction''. Once at the centre, they would form a bound pair, a binary supermassive black hole, separated by 
about 1 light year.

\begin{figure}
\epsscale{0.9}
\plotone{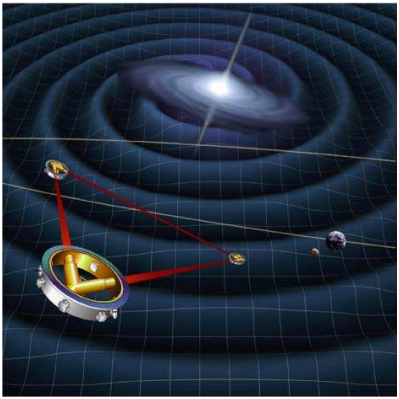}
\caption{{\bf Ripples in space-time.}
An artist's conception of the Laser Interferometer Space Antenna (LISA), a proposed observatory for 
detecting gravitational radiation. LISA will consist of three satellites that will fly 5 million
km apart in an equilateral triangle  and some 50 million km away from Earth.  The passage 
of gravitational waves, like those by a pair of coalescing supermassive black holes, through 
the antenna would cause the distances between the satellites to oscillate very slightly.  This 
motion would be measured by laser beams between the satellites.
}
\end{figure}

Begelman, Blandford and Rees suggested that many peculiar properties of active galaxies and 
quasars might be explainable if some of these systems contained binary supermassive black holes at their centres. 
For example, some active galaxies emit radio jets that twist symmetrically on either side of the 
nucleus, suggesting that the supermassive black hole producing the jets is wobbling like a precessing top. This is 
exactly what would happen in a binary supermassive black hole -- the spinning black hole that produces the jets would 
precess as it orbits around the other black hole, just as the Earth's axis wobbles due to the 
gravitational pull of the Sun and the Moon.

Other active galaxies show periodic shifts in the amplitude or Doppler shift of their emission. The 
best-studied case, a quasar called OJ 287, has experienced several major outbursts every 12 
years since monitoring began in 1895. These flares could be produced by a small supermassive black hole 
($10^8$ solar masses) passing through the accretion disc of a larger one
($10^9$ solar masses) once every 12 years.

In spite of these and other examples, no completely convincing case of a binary system of 
supermassive black holes has been 
found. The separation between the two black holes -- while enormous in everyday terms -- is still 
small in astronomical terms and would be difficult to observe in any but the nearest galaxies. But 
there is little doubt in the minds of astronomers that binary supermassive black holes are produced in galaxy mergers 
and that they must last for at least some period of time, some 10 million years or more, before 
coalescing into a single, even more massive black hole.

This final coalescence is a consequence of Einstein's equations, the same equations that were 
first used by Karl Schwarzschild to predict the existence of black holes. As the two black holes in a 
binary system orbit each other, they emit energy in the form of gravitational waves, ripples in 
space--time that propagate outward at the speed of light. Any accelerating mass produces this 
kind of radiation, but the only systems that can produce gravitational waves of appreciable 
amplitude are pairs of "relativistically compact" objects -- black holes or neutron stars -- in orbit 
about each other. Gravitational waves carry away energy, and so a system emitting gravitational 
radiation must lose energy -- in the case of a binary black hole, this means that the two black holes 
must spiral in toward each other. The infall would be slow at first, but would accelerate until the 
final plunge when the two black holes coalesced into a single object.

\bigskip
\noindent
{\bf Gravitational-wave signatures}

\noindent
The coalescence of a binary supermassive black hole would be one of the most energetic events 
in the universe. However, virtually all of the energy would be released in the form of gravitational 
waves, which are extremely difficult to detect; there would be little if any of the electromagnetic 
radiation (light, heat, etc.) that make supernova explosions or quasars so spectacular.

No direct detection of gravitational radiation has ever been achieved. But the prospect of detecting 
gravitational waves from coalescing black holes is extremely exciting to physicists: it would 
constitute robust proof of the existence of black holes and it would permit the first real test of 
Einstein's relativity equations in the so-called strong-field limit. Furthermore by comparing the 
gravitational waves of coalescing black holes with detailed numerical simulations, the masses, 
spins, orientations and even distances of the two black holes could in principle be derived.

The prospect of observing the coalescence of a binary supermassive black hole is one of the primary motivations 
behind the Laser Interferometer Space Antenna (LISA), a space-based gravitational-wave 
telescope that is expected to be launched in about 2012. The ground-based gravitational-wave 
detectors that are already in operation are unable to detect the long-wavelength gravitational 
waves produced by binary supermassive black holes (see {\it Physics World} December 2001 pp10--11.)

LISA will consist of three spacecraft separated by 5 million kilometres flying in an equilateral 
triangle (figure 3). A passing gravitational wave would stretch and squeeze the 
space between the spacecraft, causing very slight shifts in their separations. Although such shifts 
are tiny -- some 10$^{-12}$ m across -- they could be detected by laser interferometers.

LISA's designers must address one important question: how frequently will the instrument 
detect a signal from coalescing black holes? LISA will have the sensitivity to detect mergers of 
supermassive black holes out to incredible distances, essentially to the edge of the observable universe. One way to 
estimate the event rate is to calculate how frequently galaxies merge within this enormous volume. 
On this basis, LISA should detect at least one event every few years. However the situation is 
more complicated than this, since it is only the final stages of black-hole coalescence that produce 
an observable signal. At this stage, the separation between the two black holes has fallen below 
about 0.01 light years.

In order to reach such small distances, the black holes must first spiral together from their initial 
separation of about 1 light year. Gravitational radiation itself is too inefficient to achieve this; 
some other mechanism must first extract energy from the binary or else the decay will stall at a 
separation too great to generate a measurable signal for gravitational wave telescopes.

\begin{figure}
\epsscale{1.0}
\plotone{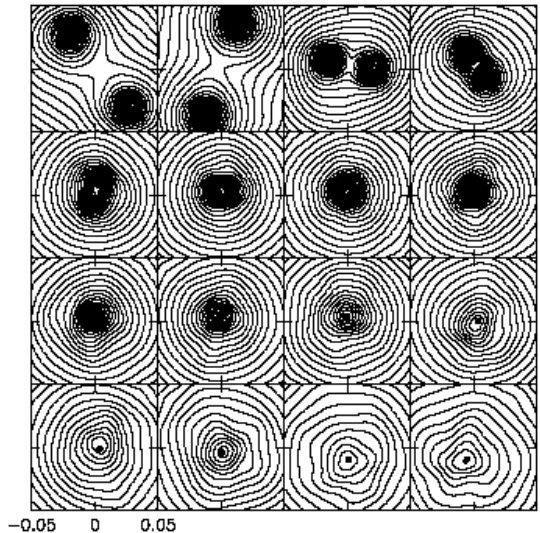}
\caption{{\bf Simulations follow galaxy union.}
A supercomputer model of two galaxies merging.
The contours indicate the density of stars around the black holes, 
while the black dots represent the black holes.
At the end of the merger (eighth and ninth panels), the black holes are separated by
about 1 light-year and form a binary system. This massive system then ejects stars from the nucleus, lowering 
the density. Eventually the two black holes would coalesce after emitting a powerful burst of 
gravitational radiation.
}
\end{figure}

\bigskip
\noindent
{\bf Supermassive black holes meet supercomputers}

\noindent
To understand these issues better, our group at Rutgers is studying the formation and evolution of 
binary supermassive black holes. One of us (DM) in collaboration with Milos Milosavljevic has carried out 
supercomputer simulations of merging galaxies (figure 4). Each galaxy in the simulation initially 
contained a central point mass representing a supermassive black hole. As the galaxies come together, 
the two black holes 
fall to the centre of the merged system and form a binary in which the black holes are separated by 
about 1 light year. Once this happens, a new mechanism comes into play called the 
gravitational slingshot. Any star that passes near to the binary supermassive black hole is accelerated to high 
velocities and ejected, taking energy away from the binary and causing its orbit to decay slightly. 
As a result, the separation between the supermassive black holes gradually shrinks. We found that the gravitational 
slingshot is efficient at reducing the separation of the black holes from 1 to about 0.1 light years. 
At that point, however, the decay tends to stall since there are few stars left in the nucleus for
further interactions.

Ongoing research is investigating whether most galaxies should have uncoalesced supermassive-black-hole 
binary systems at their centres. Other processes, such as interaction of the black holes with giant gas 
clouds, might also be able to extract energy from the binary and bring the black holes closer 
together. But it seems likely that binary systems would persist in at least some galaxies, with 
interesting consequences. For instance, if a third supermassive black hole should fall into the centre 
of such a galaxy, the three objects would undergo a gravitational slingshot. The 
resulting violent interaction will eject one or more of the black holes from the nucleus, and possibly from 
the entire galaxy. In this way, rogue supermassive black holes might be created that drift forever between the galaxies.

\begin{figure}
\epsscale{2.}
\plottwo{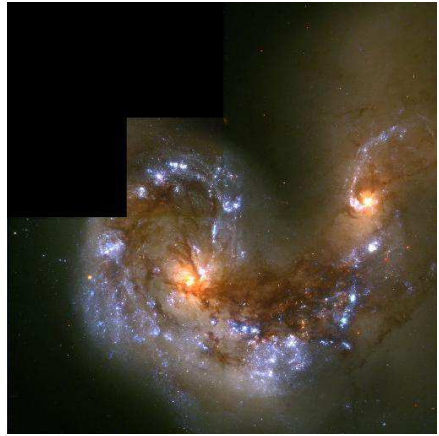}{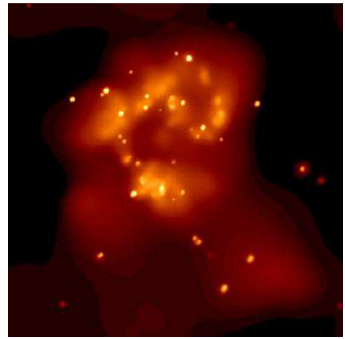}
\caption{{\bf The missing link?}
(a) The Hubble Space Telescope has revealed more than 1000 bright clusters of stars,
which were formed in a collisions between the ``antennae galaxies''
NGC 4038 and NGC 4039.
(b) The bright point-like sources in this Chandra X-ray image of the same galaxies
point to the existence of intermediate-mass black holes outside the galactic centre. 
}
\end{figure}

Most astronomers would prefer to believe supermassive black holes are confined to galactic nuclei -- if only for the 
reason that almost all galaxies in the nearby universe appear to contain one at their centres. But 
while the details of the formation and evolution of binary supermassive black holes are still being debated, 
astronomers have already begun to find support for the basic merger picture. For instance, the 
ejection of stars from a galactic nucleus by a binary system would drastically lower the density of 
stars there, whether or not the black holes finally coalesced. This prediction is in excellent agreement 
with observations: the largest elliptical galaxies, which statistically have experienced the most 
mergers, have the lowest central densities of stars.

Another prediction concerns the spin of black holes. If two supermassive black holes coalesce, their orbital motion 
during the final plunge is converted into rotation of the resulting object. This means that supermassive black holes 
at the centres of galaxies should be rotating rapidly. Furthermore, the directions of their spin axes 
should be essentially random, since the mergers responsible for imparting the spin take place from 
random directions. This last prediction has been verified by Anne Kinney and collaborators at the 
Space Telescope Science Institute in Baltimore. They showed that the orientations of radio jets in 
active galaxies -- which are thought to be in the same direction as the spin axis of the black hole -- are 
random with respect to the orientations of their host galaxies.

\bigskip
\noindent
{\bf What is still to come?}

\noindent
Looking back over the past five or six years, the progress made in our understanding of supermassive black holes has 
been respectable. But the picture is far from complete, and the road ahead is full of challenges. 
The question first raised by Fabian and Canizares more than 10 years ago is still outstanding: 
if nearby galaxies contains supermassive black holes, what prevents them from shining like quasars? By 
and large, it is not any shortage of fuel. For instance, in a giant elliptical galaxy like M87, even the 
winds from massive stars should produce a sufficient quantity of gas and dust to power the central 
supermassive black hole at a level far above what is observed.

One possible, though controversial, answer was provided by Setsuo Ichimaru of 
Tokyo University back in 1977 -- albeit in the different context of accretion onto 
stellar mass black holes. Ichimaru's idea was revived by Rees, Sterl Phinney of Caltech, 
Begelman and Blandford in 1982: the material accreting onto the central supermassive black hole might be organized 
in ``a torus of gas too hot and tenuous to radiate efficiently.''  
Later, in the 1990s, Ramesh Narayan at Harvard University developed 
a rigorous mathematical formalism for advection dominated accretion flows (ADAFs) -- flows in 
which most of the energy liberated by the accreting gas is ``advected'' inward, rather than radiated 
outward. While discrepancies between the models and observations remain, there is a respectable 
and growing body of evidence in favour of ADAFs around at least some supermassive black holes.

The existence of a supermassive black hole in the nucleus of every galaxy has become the current paradigm. 
However, all of the black holes detected so far have masses above about 1 million solar masses, while black 
holes created in supernovae explosions are much smaller, up to 15 solar masses. Essentially 
nothing definite is known about the existence of ``intermediate mass black holes'' that are
between 100 and 1 million times more massive than the Sun. We recently analysed Hubble 
data of the faint nearby galaxy M33 and found that the mass of any black hole at its centre 
could not exceed about 3000 solar masses (see Merritt {\it et al.} in further reading). Only an upper limit 
could be given because the sphere of influence was too small to be resolved 
even by the Hubble telescope. As frustrating as this might seem, this upper limit is the closest we 
have come to demonstrating the existence of intermediate mass black holes.

Resolving the issue of whether these objects exist goes beyond mere bookkeeping: it might be crucial for 
understanding how supermassive black holes form. 
Recent images of M82 and the ``antennae'' galaxies, taken with the 
Chandra satellite, suggest that intermediate mass black holes might reside in the 
cores of star clusters outside the galactic 
centre (figure 5). With time, such star clusters might 
spiral towards the centre of their host galaxy -- thanks to the same dynamical friction process 
already described. 
The intermediate mass black holes deposited at the galaxy's centre might subsequently merge to form 
supermassive ones. In other words, Chandra's peripheral IBHs might be the building blocks of Hubble's central 
supermassive black holes.

But radically different scenarios have also been proposed. At the opposite extreme, supermassive black holes might 
have been formed very rapidly and very early on in the universe's history, when their host galaxies 
looked nothing like they do today. During the very early stages of structure formation, local 
perturbations might have led to the growth of dark matter halos,
which might have catalysed the formation of supermassive black holes at their centre. 
Ironically, in this scenario very massive black holes form more easily
than small ones; in fact all black holes are predicted to exceed one million solar 
masses. The subsequent growth of the black hole could then control the formation and appearance of 
the galaxy around it.

While we might not yet know the full story about supermassive black holes,
there is at least one certainty -- the next few years will be very interesting indeed.

\clearpage

{\bf Laura Ferrarese} and {\bf David Merritt} are in the Department of Physics and Astronomy, 
Rutgers University, 136 Frelinghuysen Road, Piscataway, NJ 08854 8019, e-mail 
lff@physics.rutgers.edu and merritt@physics.rutgers.edu 

\bigskip\bigskip

\noindent
{\bf Further reading}

\noindent
M Bartusiak 2000 {\it Einstein's Unfinished Symphony: Listening to the Sounds 
of Space-Time} (Joseph Henry Press, Washington DC)

\noindent
L Ferrarese and D Merritt 2000 A fundamental relation between supermassive 
black holes and  their host galaxies {\it Astrophys. Jour.} {\bf 539} L9

\noindent
J Kormendy and D Richstone 1995 Inward bound --the search for supermassive 
black holes in galactic nuclei {\it Ann. Rev. Astr. and Astrophys.} {\bf 33} 
581

\noindent
D Merritt, L Ferrarese and C L Joseph 2001 No supermassive black hole in 
M33? {\it Science} {\bf 293} 1116

\noindent
M Milosavljevic and D Merritt 2001 Formation of galactic nuclei 
{\it Astrophys. Jour.} {\bf 563} 34

\noindent
K S Thorne 1994 {\it Black Holes and Time Warps: Einstein's Outrageous Legacy}
(WW Norton and Company, New York)

\clearpage 

\end{document}